\pdfoutput=1 
\documentclass{JINST}
\let\ifpdf\relax
\usepackage{textcomp} 
\usepackage{amsmath}
\newcommand*\diff{\mathop{}\!\mathrm{d}}
\usepackage{float}
\usepackage[font=small,format=plain,margin=0.1\textwidth]{caption}

\title{The Highly Miniaturised Radiation Monitor}

\author{E.~F.~Mitchell$^a$\thanks{Corresponding author.}, H.~M.~Ara{\'u}jo$^a$, E.~Daly$^b$, N.~Guerrini$^c$, S. Gunes-Lasnet$^d$, D.~Griffin$^d$, A.~Marshall$^d$, A.~Menicucci$^b$, T.~Morse$^d$, O.~Poyntz-Wright$^d$, R.~Turchetta$^c$, S.~Woodward$^d$\\
\llap{$^a$}High Energy Physics Group, Department of Physics, Imperial College London,\\
Prince Consort Road, London SW7 2AZ, UK.\\
\llap{$^b$}Space Environments and Effects Section, European Space Research and Technology Centre,
European Space Agency, Keplerlaan 1, 2200 AG Noordwijk, The Netherlands.\\
\llap{$^c$}RAL Technology, STFC Rutherford Appleton Laboratory, Didcot OX11 0QX, UK.\\
\llap{$^d$}RAL Space, STFC Rutherford Appleton Laboratory, Didcot OX11 0QX, UK.\\
E-mail: \email{e.mitchell09@imperial.ac.uk}}

\abstract{We present the design and preliminary calibration results of a novel highly miniaturised particle radiation monitor (HMRM) for spacecraft use. The HMRM device comprises a telescopic configuration of active pixel sensors enclosed in a titanium shield, with an estimated total mass of 52~g and volume of 15~cm$^3$. The monitor is intended to provide real-time dosimetry and identification of energetic charged particles in fluxes of up to 10$^8$~cm$^{-2}$~s$^{-1}$ (omnidirectional). Achieving this capability with such a small instrument could open new prospects for radiation detection in space.}

\begin{document}

\section{Introduction}\label{intro_sec}

In this article we describe the development of the Highly Miniaturised Radiation Monitor (HMRM), an innovative particle detector intended for spacecraft radiation environment monitoring. The HMRM comprises a telescopic configuration of application-specific integrated circuit (ASIC) active pixel sensors enclosed in a titanium shield. It is intended to provide real-time dosimetry and identification of energetic charged particles in fluxes of up to 10$^8$~cm$^{-2}$~s$^{-1}$ (omnidirectional).

The particle radiation environment in Earth orbit is dominated by protons (energies up to $\sim$500~MeV) and electrons (energies up to $\sim$10~MeV) trapped in the Earth's magnetosphere. Smaller particle fluxes of solar and galactic origin, which may include ion species, are also present. Interactions of these particles with spacecraft systems and payloads lead to the cumulative degradation of components and to single-event errors in electronic devices. Safety concerns for human space habitation and exploration pose even greater challenges from this perspective. The data provided by monitoring devices help in assessing these risks and in correlating radiation effects with the radiation environment. This may result in improved mission planning, recommendations for spacecraft design and introduces the possibility of real-time alerting.

In contrast to scientific payloads, small support instruments often have limited functionality (e.g.~simple dosimetry) and offer little or no particle discrimination. Damage effects depend strongly on particle species and energy, meaning that particle identification would be an important advantage for these devices. The development of a small, accurate instrument suitable for widespread use on satellites in any orbit could therefore open new prospects for radiation detection in space.

This paper provides a summary of the design of the HMRM, together with initial results of the ASIC sensor calibration. Section~\ref{geom_sec} presents the particle detector geometry, optimised through the use of Monte Carlo simulations. Section~\ref{sensor_sec} describes the design of the pixel sensor ASICs, which is also discussed in~\cite{guerrini2013}. The algorithms used to categorise particles and to reconstruct incident particle energy spectra are discussed in Section~\ref{algo_sec}. Finally, Section~\ref{cal_sec} presents preliminary results of the sensor calibration.

\section{Monitor design}\label{geom_sec}

Particle identification ability depends both on the chosen sensor technology and on the overall identification `scheme'. This scheme encompasses the sensor arrangement, monitor geometry and materials selection, together with the analysis algorithm used to convert sensor data into an interpretation of the radiation environment. All of these were considered when developing the HMRM architecture and when optimising its design with respect to its detection performance.

\subsection{Conceptual design}~\label{cmodes}
The HMRM conceptual design comprises a telescopic stack of multiple sensors through which particles may pass. This, together with the additional control electronics, is surrounded by a casing which provides structural support and shielding of low energy particles. An aperture allows restricted admission of these lower energy particles, while a thin aperture cover blocks visible and ultraviolet photons (to which the sensors are also sensitive). Figure 1 illustrates the HMRM implementation adopted after taking into account the following considerations, which are more generic in nature.

Particles are identified, where possible, from the characteristic combination of energy deposits in the sensors. The mean energy loss rate in each sensor, dE\textfractionsolidus dx, can be estimated from these deposit measurements, allowing the particle identity to be deduced. Although different particles with similar velocities may have equal energy loss rates, sampling over multiple sensors reveals differences in behaviour. This is achieved by slowing the particle with inter-sensor material so that each dE\textfractionsolidus dx value is sampled at a different velocity. The energy deposit variance arising from the fundamentally stochastic energy loss processes, together with the loss of particles scattering out of the sensor telescope, leads to statistical variance in individual particles. Sampling of the radiation environment with a sufficiently large number of particles can, however, effectively mitigate these effects for practical applications.

Particular relevance is attributed to the number of sensors hit by each particle as this indicates a minimum energy (dependent on the species) required to penetrate the sensor stack. In addition to measuring the sensor energy deposits, each event can be categorised with a coincidence or \textit{detector} mode, labelled C1, C2, etc., according to: 
\begin{eqnarray}
\text{C}1 &=& \text{S}_1 \times \overline{\text{S}_2} \times \overline{\text{S}_3} \times \overline{\text{S}_4} \times \mathellipsis\\
\text{C}2 &=& \text{S}_1 \times \text{S}_2 \times \overline{\text{S}_3} \times \overline{\text{S}_4} \times \mathellipsis\\
\text{C}n &=& \prod_{i=1}^n \text{S}_{i} \times \prod_{i=n+1}^N \overline{\text{S}_{i}},
\end{eqnarray}
where $\text{S}_n$ indicates a detected hit to the $n^{th}$ closest sensor to the entrance window and $\overline{\text{S}_n}$ indicates no hit within a single sensor integration period. This labelling system can be extended to a telescope comprising any number of sensors, resulting in \textit{N} overall modes for \textit{N} sensors.

\begin{center}
\begin{figure}[h]
\begin{center}
\includegraphics[width=1\textwidth]{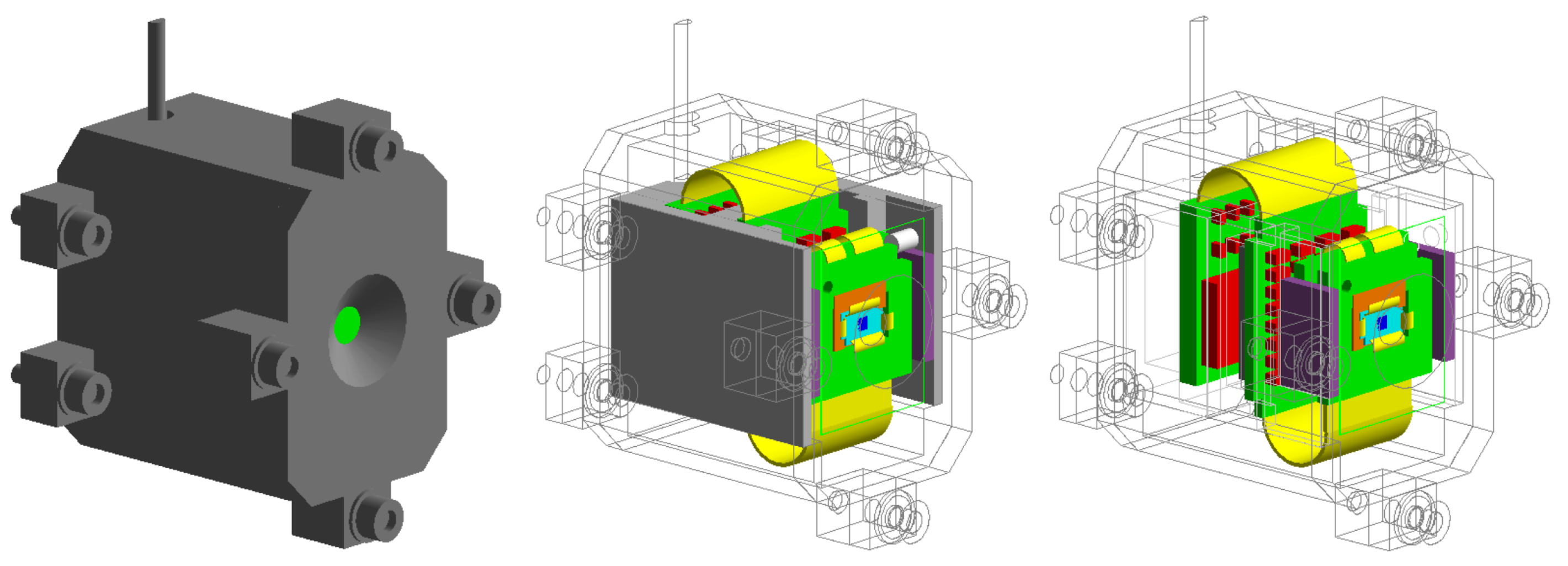}
\caption{The HMRM version 1.0 Geant4 geometry model, shown in three views revealing the internal structures. Left to right: (1) the titanium casing, aperture and aluminium window; (2) internal support, front PCB, first sensor; (3) four PCBs (electronic components are approximate only), flex harnesses. The outer dimensions of the casing, excluding bolts/lugs, measure 1.7~cm~\texttimes~2.4~cm~\texttimes~2.2~cm.}
\label{full_geom}
\end{center}
\end{figure}
\end{center}

\subsection{Optimised geometry}

Various parameters of the detector geometry were identified which, together with a selection of figures of merit and design principles, were used to guide the geometry optimisation. A series of incremental changes to the baseline conceptual design were evaluated by simulating simplified models with the Geant4 Monte Carlo toolkit~\cite{G4_agostinelli,G4_allison}. The optimised characteristics were incorporated into the full monitor prototype design, which features a stack of four application-specific CMOS active pixel sensors controlled by an FPGA device (Actel Igloo AGL1000). 

Each sensor comprises a 50\texttimes 50 pixel array, providing a 1~mm$^2$ area and 12~\textmu m sensitive (epitaxial layer) thickness. A minimum-ionising particle at normal incidence is expected to provide a most-probable signal of $\sim$1000~e$^-$, while single pixel noise is expected to be of order 10~e$^-$~rms. Ionisation charge is generally collected by multiple adjacent pixels; after digitisation the pixel signals are summed across each sensor, resulting in a set of four charge sums. These data are used to derive radiation data, as explained later in Section~\ref{algo_sec}. Single particle detection is achieved by using a short integration time of 12.5~--~100~\textmu s which is adaptively controlled with an electronic shuttering algorithm. This, together with the restricted geometric acceptance, ensures that pile-up (two or more simultaneous hits) is expected in fewer than 5$\%$ of exposures at peak omnidirectional flux. 

In addition to sensor readout, the FPGA provides data processing, telemetry and communications via a CAN bus, and other support functions. The full monitor design was implemented as the detailed simulation geometry shown in Fig.~\ref{full_geom}.

\begin{center}
\begin{figure}[h]
\begin{center}
\includegraphics[width=0.55\textwidth]{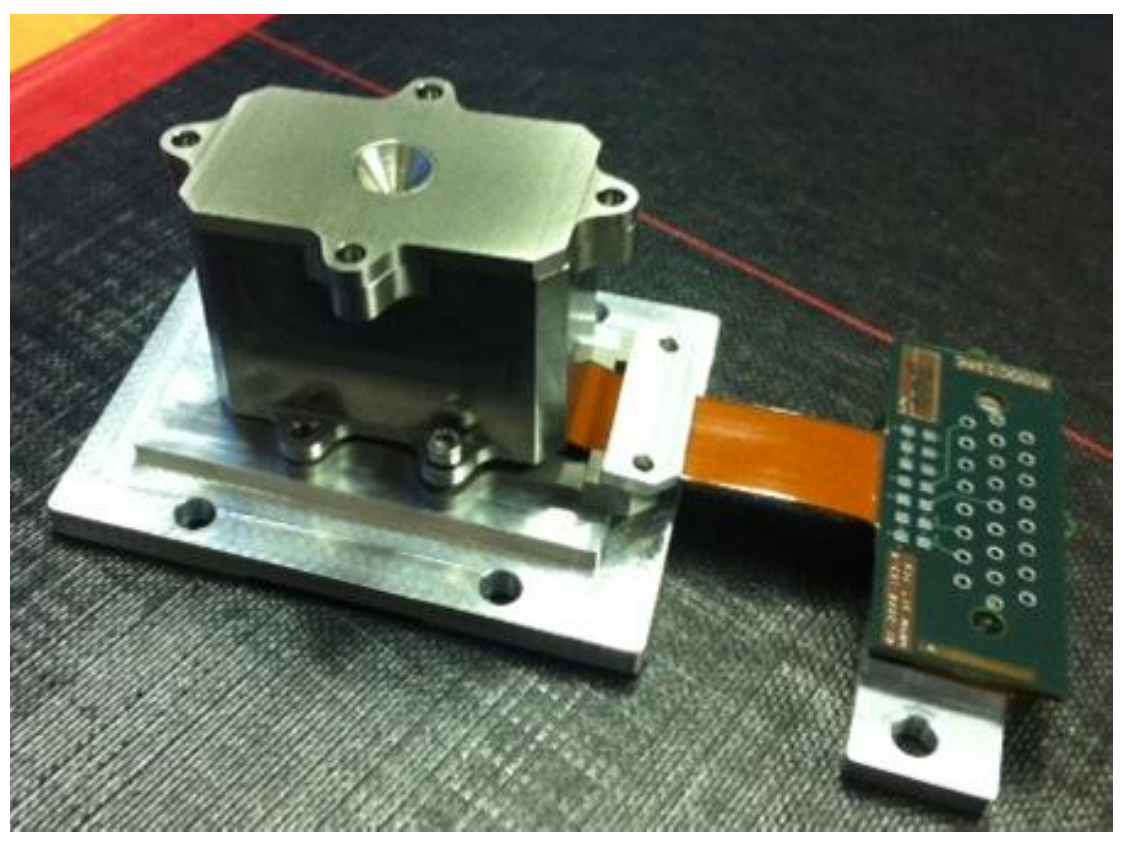}
\caption{Photograph of the assembled HMRM.}
\label{photo}
\end{center}
\end{figure}
\end{center}

\begin{figure}[ht]
\begin{minipage}[b]{0.5\linewidth}
\centering
\includegraphics[width=\textwidth]{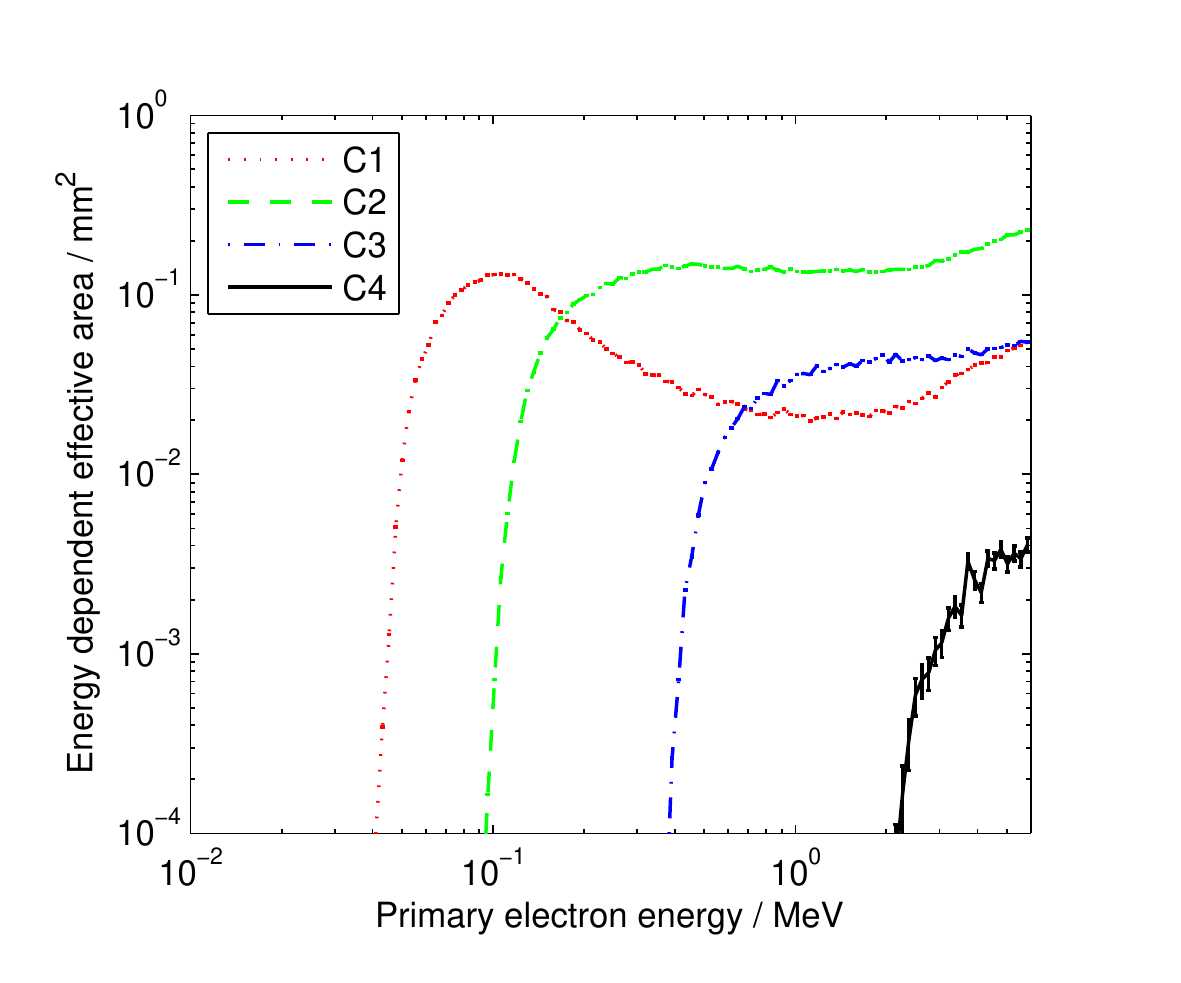}
\end{minipage}
\hspace{0.5cm}
\begin{minipage}[b]{0.5\linewidth}
\centering
\includegraphics[width=\textwidth]{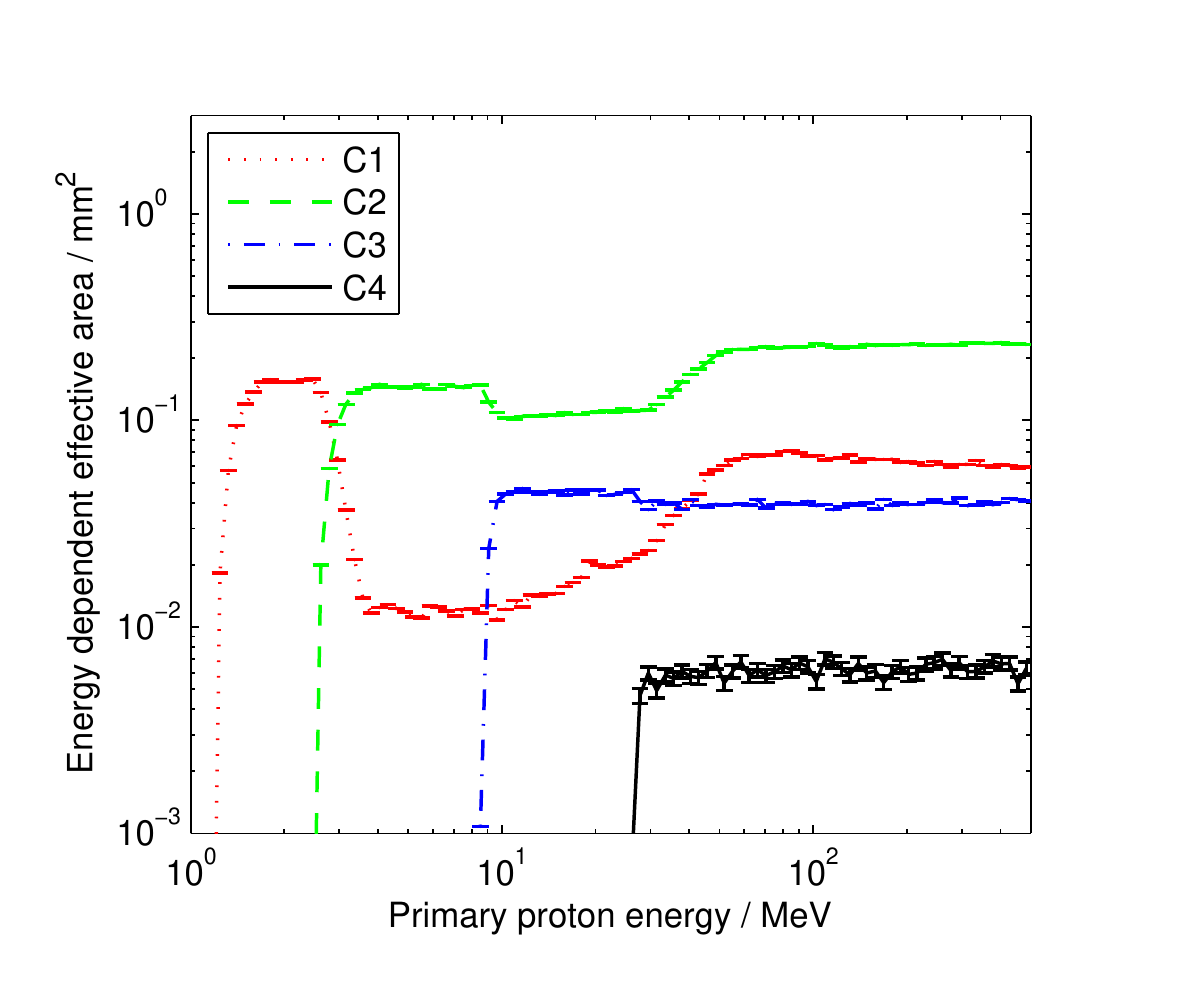}
\end{minipage}
\caption{Energy dependent effective area for the four detection modes, C1--4, for isotropic electrons (left) and protons (right).}
\label{area_plot}
\end{figure}

The prototype HMRM assembly is shown in Fig.~\ref{photo}. The geometric acceptance of the monitor was assessed through extensive simulations and is expressed as an effective area (omnidirectional quantity) for electron and proton acceptance in Fig.~\ref{area_plot}. Further simulations were used to assess the monitor's suitability in a series of reference Earth orbits~\cite{EMthesis}.

\section{ASIC sensor design}\label{sensor_sec}

Application-specific CMOS Active Pixel Sensors (APS) were developed for use in the HMRM project. This technology, having already found widespread use for visible light imaging~\cite{mendis_94}, is in development for particle physics applications (e.g.~electromagnetic calorimetry and tracking~\cite{stanitski_2011}). The use of a CMOS fabrication process allows the signal analysis circuitry to be included on-chip, and within each pixel, resulting in a compact, rugged device with very low noise. Suitability for space applications is demonstrated by the existing use of radiation-hardened CMOS MAPS as space-based imagers~\cite{sayed_01} and star trackers~\cite{liebe_02}. These use enclosed geometry transistors and guard rings to reduce radiation-induced leakage currents, giving maximum operational doses of up to $\sim$300~kGy.

A commercially-available 0.18~\textmu m CMOS Image Sensor technology was selected to reduce development cost. Each HMRM ASIC incorporates a 50\texttimes50 APS array with a pixel pitch of 20~\textmu m (total sensitive area of 1~mm$^{2}$). All pixels are read out simultaneously (`snapshot mode') and digitised via a 3-bit column parallel ADC with correlated double sampling (CDS). Each ADC comparator level is programmed as a 7-bit threshold setting, allowing customisable, non-linear pixel digitisation schemes. This process achieves a maximum frame rate of $\sim$10~kHz.

\begin{center}
\begin{figure}[h]
\begin{center}
\includegraphics[width=0.8\textwidth]{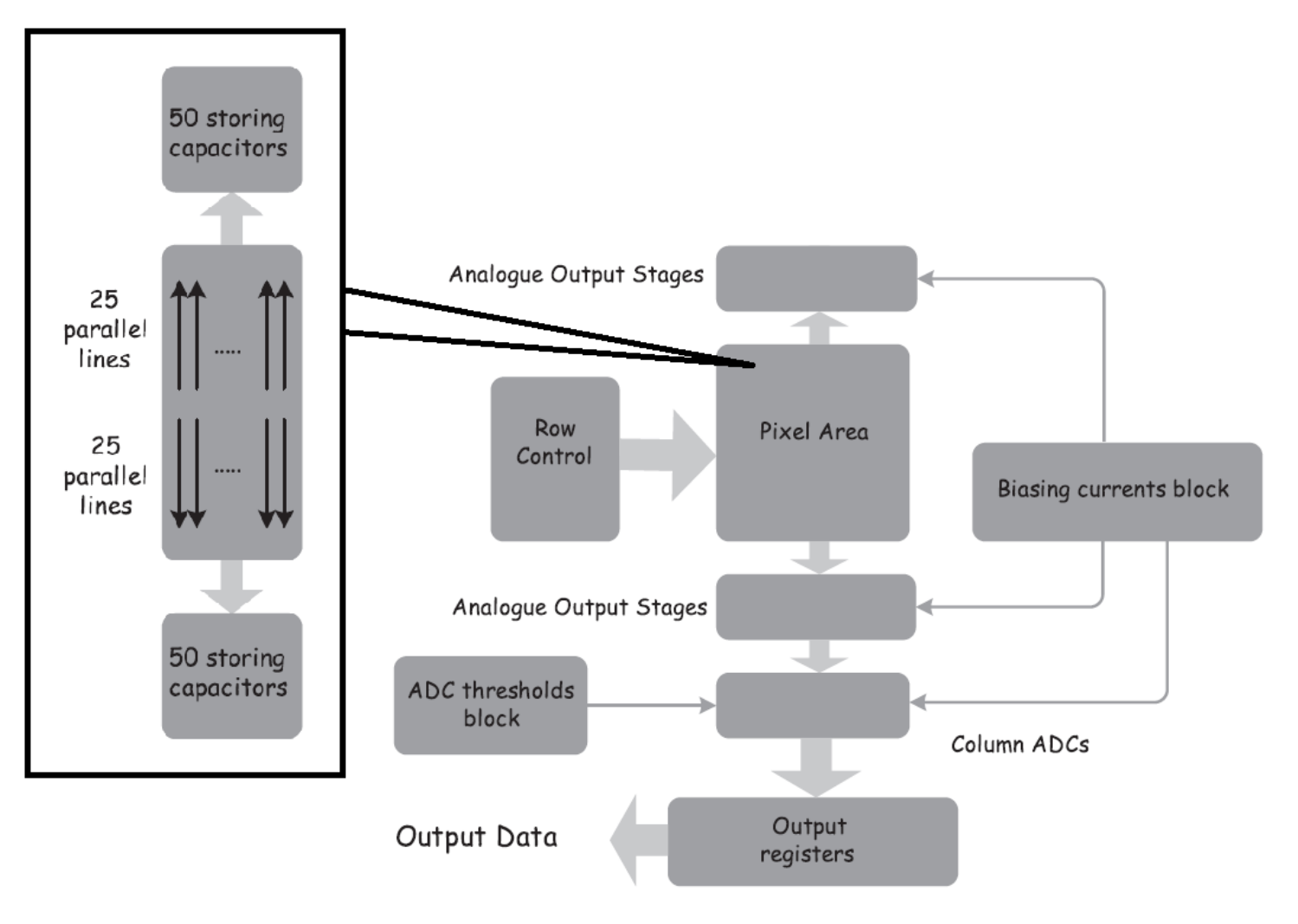}
\caption{HMRM ASIC readout block diagram.}
\label{ASIC_flowchart}
\end{center}
\end{figure}
\end{center}

A secondary analogue output channel was also implemented to allow initial debugging of
the ASIC. To reduce the number of external components and the complexity of the PCB, biasing currents and ADC thresholds are generated by on-chip DACs controlled by externally programmable registers. Band-gap voltage references and a 9-bit temperature sensor are also included. A high-level block diagram of the readout chain is given in Fig.~\ref{ASIC_flowchart}.

The selected pixel architecture uses four transistors (4T), reducing noise while maintaining good radiation hardness.
The noise performance of a similar 4T pixel already fabricated by RAL-STFC has been characterised at total ionising doses up to 110~kRad. This investigation found a $\sim$40$\%$ increase in noise and a 450$\%$ increase in dark current as the dose was increased from 1~kRad to 100~kRad~\cite{coath2010}. As the total dose requirements of the HMRM were set at 100--150~kRad, a modest noise increase is anticipated but is considered acceptable for this specific application. In addition, the HMRM has the capability of masking individual pixels with undesirably high noise.

\subsection{Readout and digitisation}

All pixels are read simultaneously and transferred outside of the array, where both the reset and the integrated values are stored on capacitor banks. Data from the storage capacitors are multiplexed to the input of each ADC, one of which is provided for each half-column of 25 pixels. Targeting a frame rate of 10,000~fps, the analogue-to-digital conversion must take place in 100~\textmu s, leaving 4~\textmu s per pixel sample.

\begin{center}
\begin{figure}[h]
\begin{center}
\includegraphics[width=0.7\textwidth]{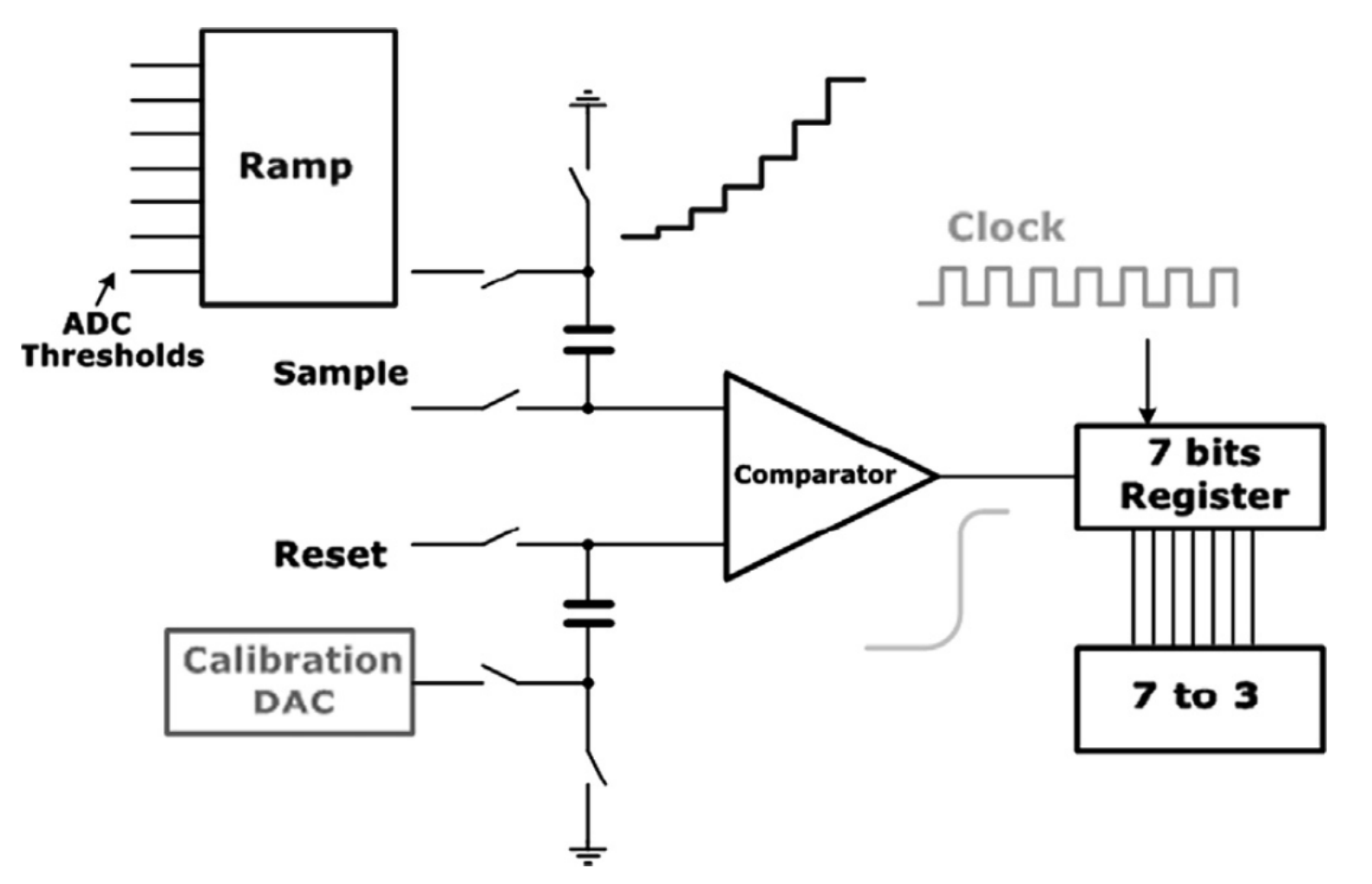}
\caption{Overview of the ADC design for the HMRM ASIC. One comparator is provided for each half-column of 25 pixels.}
\label{ADC}
\end{center}
\end{figure}
\end{center}

In designing the ADC, conflicting requirements were balanced. Low power and reliability would point towards a single-ramp ADC whereas the short conversion time and fine resolution required by the particle identification algorithm suggest a flash ADC. These considerations, and the requirement of CDS, led to the ADC architecture presented in Fig.~\ref{ADC}, which can be regarded as a hybrid between ramp and flash ADC. Reset and sample values are transferred onto two separate capacitances connected to the input of a comparator. The other plate of the capacitor, storing the reset value, is connected to an 8-bit trimming DAC that compensates for the comparator offset. Each ADC has its own trimming register, hence 816 trimming bits must be programmed before acquiring data. In a conventional flash ADC, the data to be converted is compared with $2^{N}-1$ thresholds in order to obtain $N$-bit outputs. Our solution was to use just one comparator and add the thresholds to the data; after each threshold addition the output of the comparator is stored in a 7-bit register. 

Assuming that the comparator offset has been completely cancelled by the trimming DAC, the comparator determines whether the sample (minus the reset value) is less than the threshold. The resulting seven bits are encoded into a 3-bit number per pixel and then output on a 9-bit bus. All transistors have been designed with an enclosed geometry layout to mitigate the effects of radiation. For the same reason all digital blocks are based on Triple Majority Voting (TMV).

\section{Particle detection algorithms}\label{algo_sec}

The purpose of the HMRM is to provide real-time measurements of the local particle flux. Information generated by the monitor is transmitted as data packets, which may be used by the host spacecraft to react to changes in the particle environment. Monitor output data may also be used, over longer time periods, in more detailed analyses. These are not intended to be executed within the monitor or spacecraft but rather on the Earth, as part of the ground segment, and include statistical methods to reconstruct the incident particle energy spectra.

\subsection{Data products}

Real-time data products provided by the monitor include particle count rate, ionising dose rate, cumulative dose and identified particle counts. The calculations necessary for these quantities are executed in the FPGA --- this also regulates the exposure time for which charge is integrated in the sensor pixels during each readout cycle. This shuttering algorithm has the effect of extending the count rate dynamic range: exposure time is reduced in high fluxes, reducing the probability of event pile-up. 

The information obtained in each readout cycle, comprising the four sensor energy deposit values, must be used to make a particle identification/event characterisation. This is distributed into thirty-two categories, or \textit{channels}, each with a separate 23-bit event counter. Each count corresponds to a set of four sensor measurements and thus a typical particle signature. The requirement for the monitor to have low data bandwidth led to the choice of 32~channels, although as a general method it is also scalable to any number of sensors and any number of histogram channels.

Some of the 32 channels --- those permitting high-purity particle identification --- can be used to give coarse information to the spacecraft in real time regarding the occurrence and severity of radiation events. In the present implementation, three channels are selected for this process. The remainder of the thirty-two channel counts are used for energy spectrum reconstruction in the ground segment. The output data are divided between two telemetry packet types, standard and extended, both of which are available at a preset frequency. The standard packet is just 48~bytes long and is intended for immediate use at time intervals of order 1~s.

\subsection{Event classification and identification}

The method by which events are allocated to each channel is referred to as the \textit{event identification algorithm}. This is achieved by simultaneously histogramming the energy deposit value from each sensor; for a four sensor monitor this is a four dimensional (4-D) histogram. A separate event count is kept for each 4-D bin, which is similar to a traditional pulse height channel. This method is robust and fast enough to be executed within the 100~\textmu s limiting time for each event analysis in the HMRM FPGA. Crucially, it also preserves the coincidence information of each event. 

The success of this identification method depends on the selection of the 4-D energy deposit histogram bin edges, collectively referred to as the identification look-up table, or \textit{ID~table}. Accurate Monte Carlo simulations are invaluable in this process as they model the full energy deposit variance and thus allow assessment of the misidentification probability. However, the table may also be chosen to incorporate results from experimental tests with, for example, monoenergetic particle beams. 

\begin{table}[ht]
\caption{Part of the provisional ID~table for the HMRM (8 channels out of 32 are shown). In addition to the threshold values for each sensor energy deposit, statistics concerning the simulated particle selectivity are shown.}
\begin{center}
\begin{figure}[H]
\begin{center}
 \includegraphics[width=\textwidth]{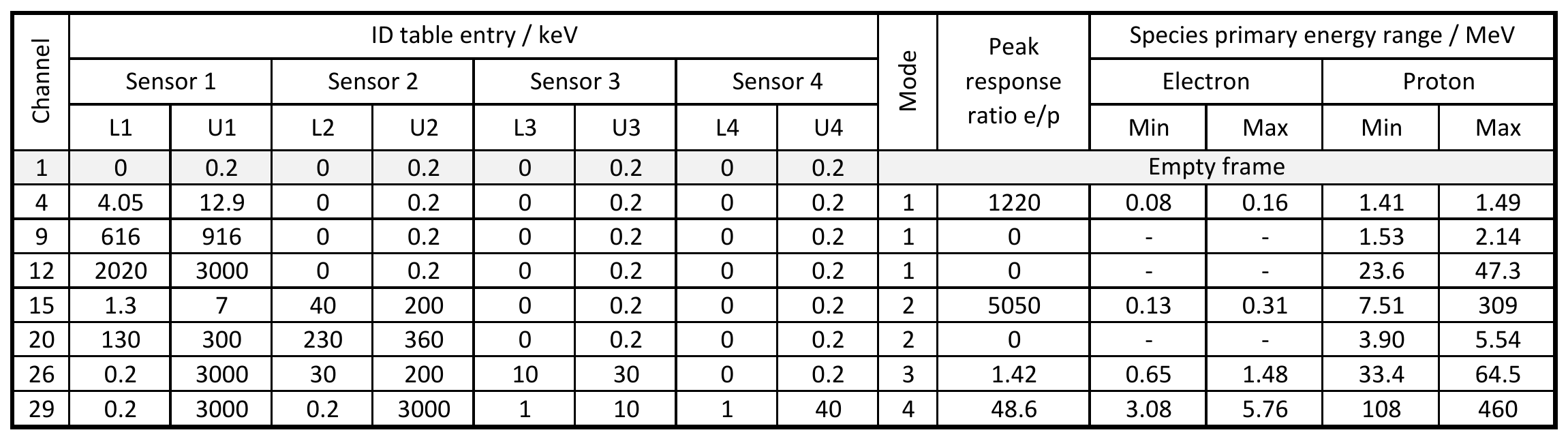}
\end{center}
\end{figure}
\end{center}
\label{ID_tab}
\end{table}

A provisional ID table has been selected, prior to experimental validation with the assembled monitor, using Monte Carlo data. Initial entries were chosen to select regions of relative particle purity in the 4-D energy deposit parameter space. These higher purity selections have generally lower detection efficiency, so mixed channels with high efficiency were also selected. While not suitable for immediate interpretation, these are important for later spectrum deconvolution. Furthermore, all channels are chosen to be mutually exclusive so that a unique allocation is possible for each event. 

Table~\ref{ID_tab} is a sample of the selected ID table and also shows the approximate characteristics of each channel. In this table the histogram bin edges for each channel are listed as a lower (L) and an upper (U) bound in energy for each sensor (L1 and U1 for sensor~1, L2 and U2 for sensor~2, etc.). Thus, for example, an event is counted in channel~15 if and only if the energy deposit in sensor~1 is in the range 1.3~--~7~keV, the deposit in sensor~2 is in the range 40~--~200~keV, and the deposits in sensors~3~and~4 are less than 0.2~keV. In this example, the deposit condition for sensors~3~and~4 is consistent with no particle hit and it is therefore a C2 mode event (see Section~\ref{cmodes}). For both species tested (electron and proton), the energy range between which the detection efficiency was greater than 50$\%$ of maximum is given, indicating the particle energy ranges selected. The peak response ratio of electrons to protons is also provided for each channel. This indicates the relative magnitudes of the selection efficiency for the two species, at their efficiency maxima. Twelve channels do not select any electrons and are thus pure proton; there are no pure electron channels, due to the greater energy deposit variance from multiple scattering, although four have a peak selection ratio greater than $10^3$ (e.g. channels 4 and 15 in Table~\ref{ID_tab}). 

\subsection{Energy spectrum reconstruction}

The energy spectra of all particle species present in an environment provide a useful description from which information about radiation effects may be derived. Given these spectra, the monitor response functions may be used to predict  
the particle counts in each of the 32 monitor channels in a given radiation flux. This  
is a simple evaluation of the formula:
\begin{equation}
\mu_k = \sum_s \int_0^T \int_{4 \pi} \int_0^{\infty} \epsilon^s_k \alpha^s f^s_u \diff E \diff \vec{\Omega} \diff t,   \label{mean_counts}
\end{equation}
where $\mu_k$ is the mean count expected in channel $k$ in an exposure time $T$. Here, the  
response functions $\epsilon^s_k$ (identification efficiency) and $\alpha^s$ (effective area) are unidirectional and  
require the unidirectional flux, $f^s_u$, to be specified. The sum is over all particle species.

Although this forward evaluation is simple, the inverse problem, where the measured counts ($N_k$) are provided  
and the $f_u^s$ must be found, is more challenging. An additional problem lies in the fact that  
the $N_k$ are integer random variables, while Equation~\ref{mean_counts} is stated in terms of the corresponding underlying  
mean values, $\mu_k$. The solution method must therefore contend with statistical fluctuation in any  
finite measurement sample.

An iterative solution method has been developed for the HMRM. Trial electron and proton trial spectra are folded with the monitor response functions to produce a set of predicted channel counts, which are compared with the actual HMRM data. The trial spectra are then modified and the process repeated until an optimal reproduction of the measured data is obtained.

This is achieved by  
assuming that each spectrum is a continuous function of particle energy and may be modelled as  
a piecewise continuous power law. $N$ free parameters are interpreted as a set of flux `nodes' at predetermined energy values.  
These are interpolated by $(N-1)$ fully-determined power law function segments to produce a trial spectrum. This spectrum is then used to modulate the response function, which may be of  
arbitrarily high resolution, to generate a set of channel count rate estimates (the `folding' process). In addition to allowing a  
greater response function resolution, it is proposed that this new model provides a better fit to  
real spectra when compared to matrix-based methods which typically assume a piecewise constant spectrum~\cite{michel_2009,ponchut_2008}.

The count rate estimates produced by the folding process are compared with monitor measurements in a least squares or maximum likelihood iterative fit. Both species must be reconstructed simultaneously since, in general, each monitor channel may count both species (mixed efficiency) and thus the two trial spectra constitute a combined hypothesis. In each iteration, evaluation of the objective function is a simple  
extension of the forward method (Equation~\ref{mean_counts}). The overall iterative process is controlled by a general purpose non-linear function minimisation routine.

Additional aspects of the spectrum reconstruction method include randomising of the energy node locations to avoid introducing bias and smoothing of the spectrum estimates produced. The latter has been achieved through a robust locally-weighted regression method~\cite{Cleveland1979}.

\begin{center}
\begin{figure}[h]
\begin{center}
\includegraphics[width=0.6\textwidth]{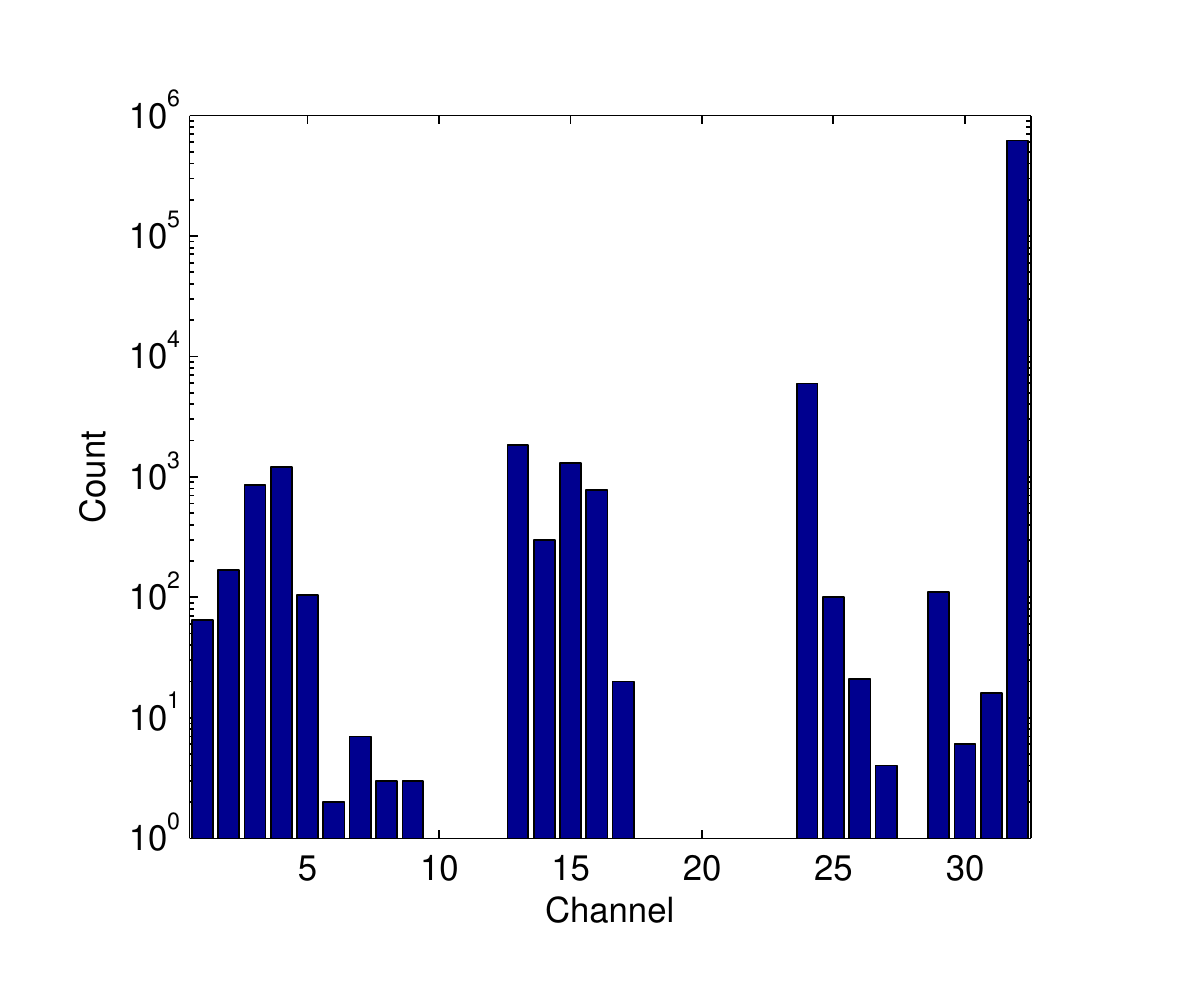}
\caption{Simulated HMRM channel counts allocated in response to a 225~s exposure in the mean radiation environment of a 23,000~km MEO.}
\label{SRcounts}
\end{center}
\end{figure}
\end{center}

\begin{center}
\begin{figure}[h]
\begin{center}
\includegraphics[trim=1cm 1.5cm 1cm 1cm, width=0.85\textwidth]{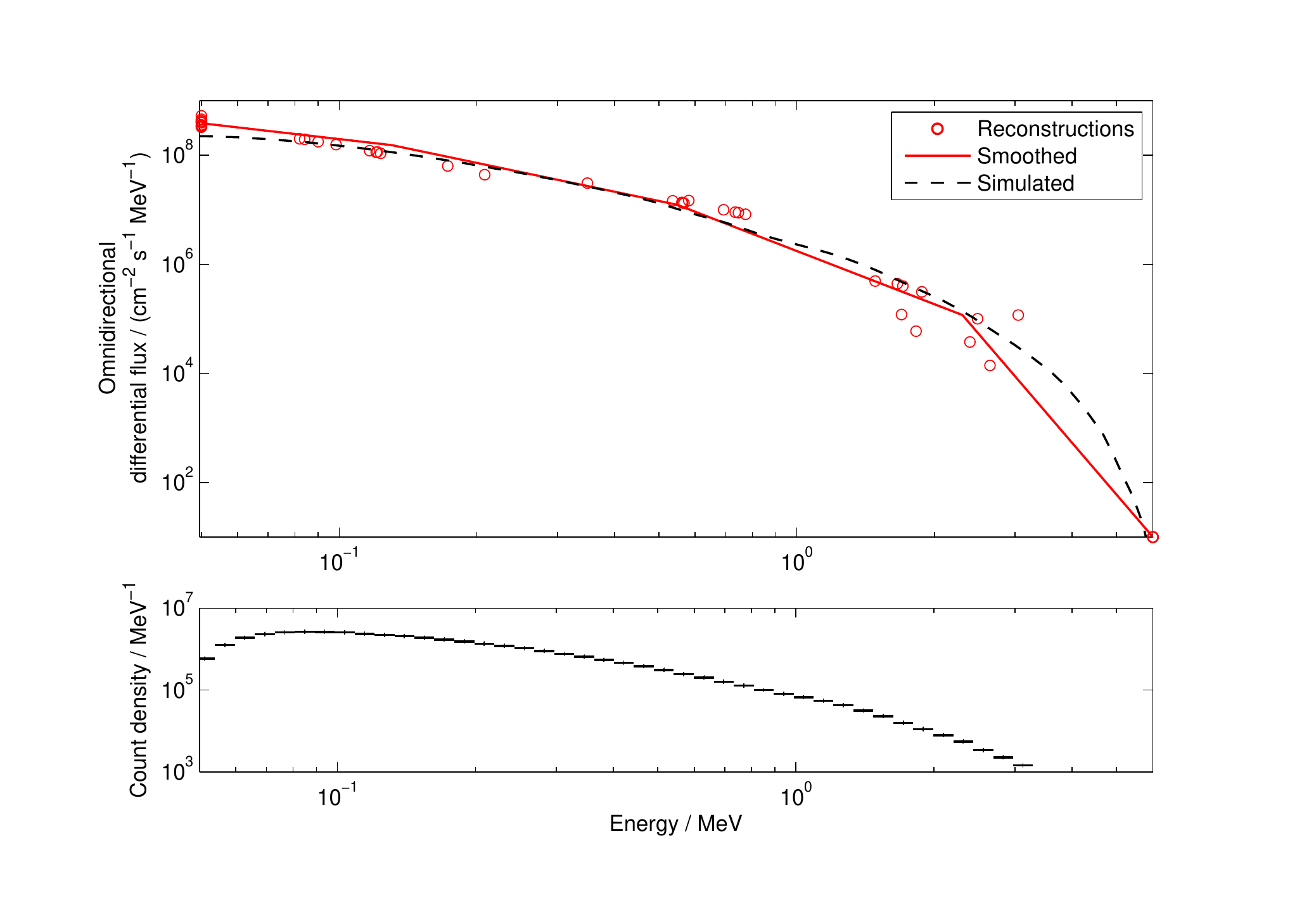}
\caption{Electron spectrum reconstruction using data from a simulated 225~s exposure in the mean radiation environment of a 23,000~km MEO. Five spectrum nodes were used; the best estimate is the result of smoothing the ten individual reconstructions.}
\label{SRspect_e}
\end{center}
\end{figure}
\end{center}

\begin{center}
\begin{figure}[h]
\begin{center}
\includegraphics[trim=1cm 1.5cm 1cm 1cm, width=0.85\textwidth]{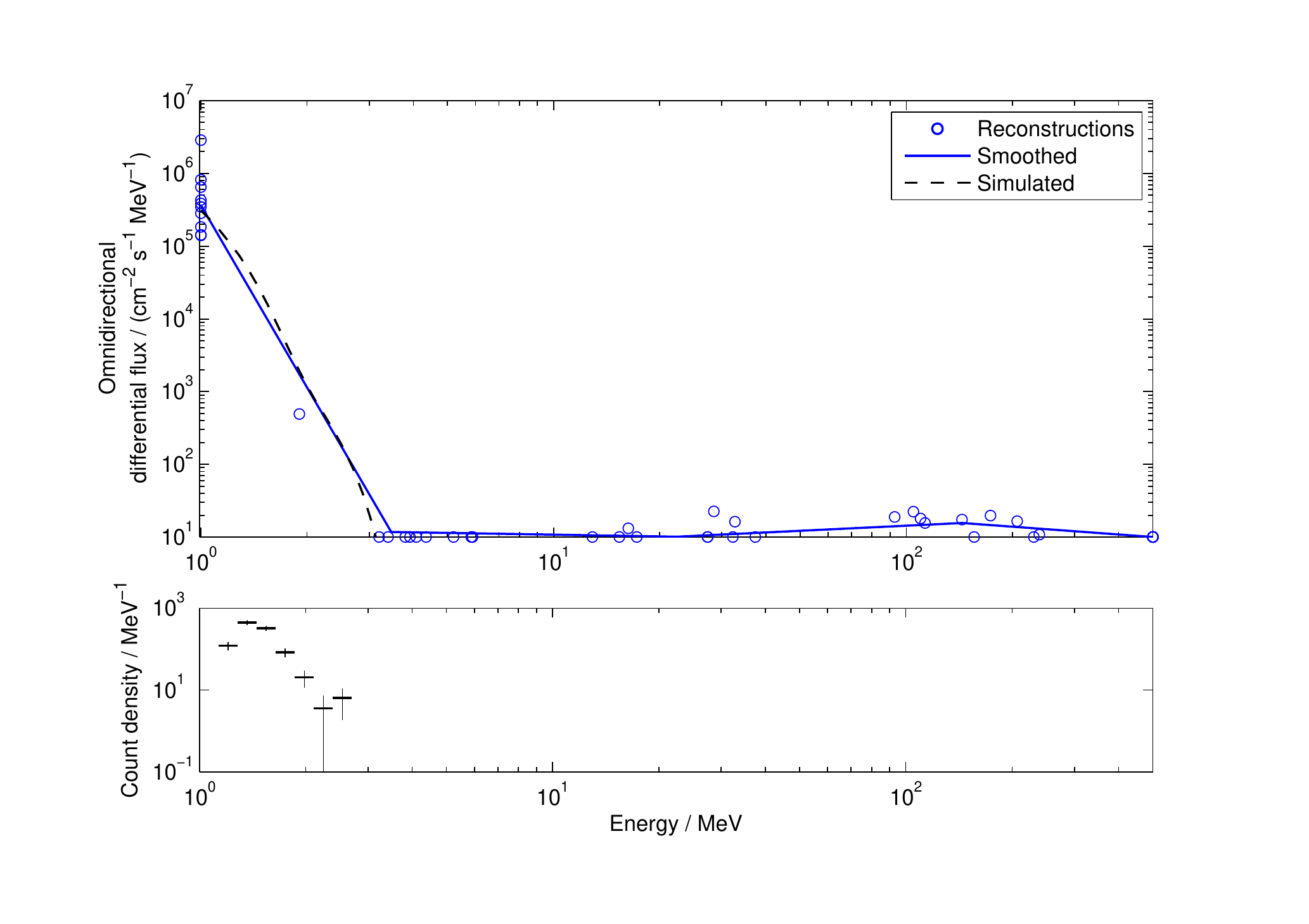}
\caption{Proton spectrum reconstruction using data from a simulated 225~s exposure in the mean radiation environment of a 23,000~km MEO. Five spectrum nodes were used; the best estimate is the result of smoothing the ten individual reconstructions.}
\label{SRspect_p}
\end{center}
\end{figure}
\end{center}

A test of the reconstruction process is shown in Figs.~\ref{SRcounts},~\ref{SRspect_e}~and~\ref{SRspect_p} using simulated data. Electron and proton mean energy spectra provided by the AE-8 and AP-8 models~\cite{ap8,ae8} were obtained for a 23,000~km altitude, 56$^\circ$ inclination medium Earth orbit (MEO) via SPENVIS~\cite{spenvis}. This orbit is typical of navigation constellation satellites and is dominated by trapped electrons with a small contribution of protons. A Geant4 simulation was run to provide a 225~s exposure in this environment, with the fluxes assumed static and isotropic over this period. A sensor response model was applied to the simulated energy deposits, involving the addition of 110~eV noise to each deposit (assuming clusters of nine pixels, each with 10~e$^{-}$ noise). The HMRM FPGA event identification algorithm was then applied to these data to produce a set of 32 channel counts, representing the modelled monitor output data in such an environment. These counts are shown in Fig.~\ref{SRcounts}. 

The reconstruction algorithm was run ten times on these counts and a smoothed best estimate of the individual results was taken for each species --- these are shown in Figs.~\ref{SRspect_e}~and~\ref{SRspect_p}. Also shown are the actual spectra simulated and the true energy distribution of the individual particles incident on the monitor sensors. Note that nominal detectable differential flux limits of 10~--~10$^{10}$~cm$^{-2}$~s$^{-1}$~MeV$^{-1}$ were enforced during the reconstruction process for both species. Five spectrum nodes were used in each reconstruction. These are directly controlled by the function minimisation routine and are interpolated to produce each trial spectrum. The use of five nodes was found qualitatively to provide a suitable compromise between spectrum resolution and under-determination of the solution.

This example reconstruction gives an excellent result; however, several simplifying assumptions have been applied. These include the restriction to two species, the neglection of pile-up and the assumption of flux isotropy. Furthermore, this reconstruction has been achieved using simulated data, where the response functions are known to accurately model the simulated monitor. Future developments will address these simplifications and attempt to verify the response functions through experimental testing.

\newpage

\vspace*{0.1cm}

\newpage

\section{ASIC calibration}\label{cal_sec}

Prior to testing of a fully assembled HMRM unit, single ASIC sensors have undergone a calibration process designed to investigate their characteristics and to determine correct operational settings. Comparator offsets, pixel pedestals and pixel noise have been measured in relative units with preliminary conversion to equivalent noise charge. While this does not constitute a complete calibration of the ASIC, these values may be used to select comparator trim settings and pixel masks for use in monitor operation.

\subsection{Pixel pedestal distribution}

The signal distribution for each pixel in one ASIC has been determined by scanning a comparator threshold level across its full range, producing the cumulative probability distribution. In general, this has the form of a sigmoid curve as the probability of the pixel signal being below threshold changes from zero for very low thresholds to one at very high thresholds. Without radiation, the distribution of dark signals (for a single pixel) is well approximated by a Gaussian. Pixel characteristics, such as the pedestal and noise (signal mean and standard deviation), have therefore been extracted by fitting an Error function to the threshold scan data of each pixel.

\begin{figure}[ht]
\begin{minipage}[b]{0.5\linewidth}
\centering
\includegraphics[width=\textwidth]{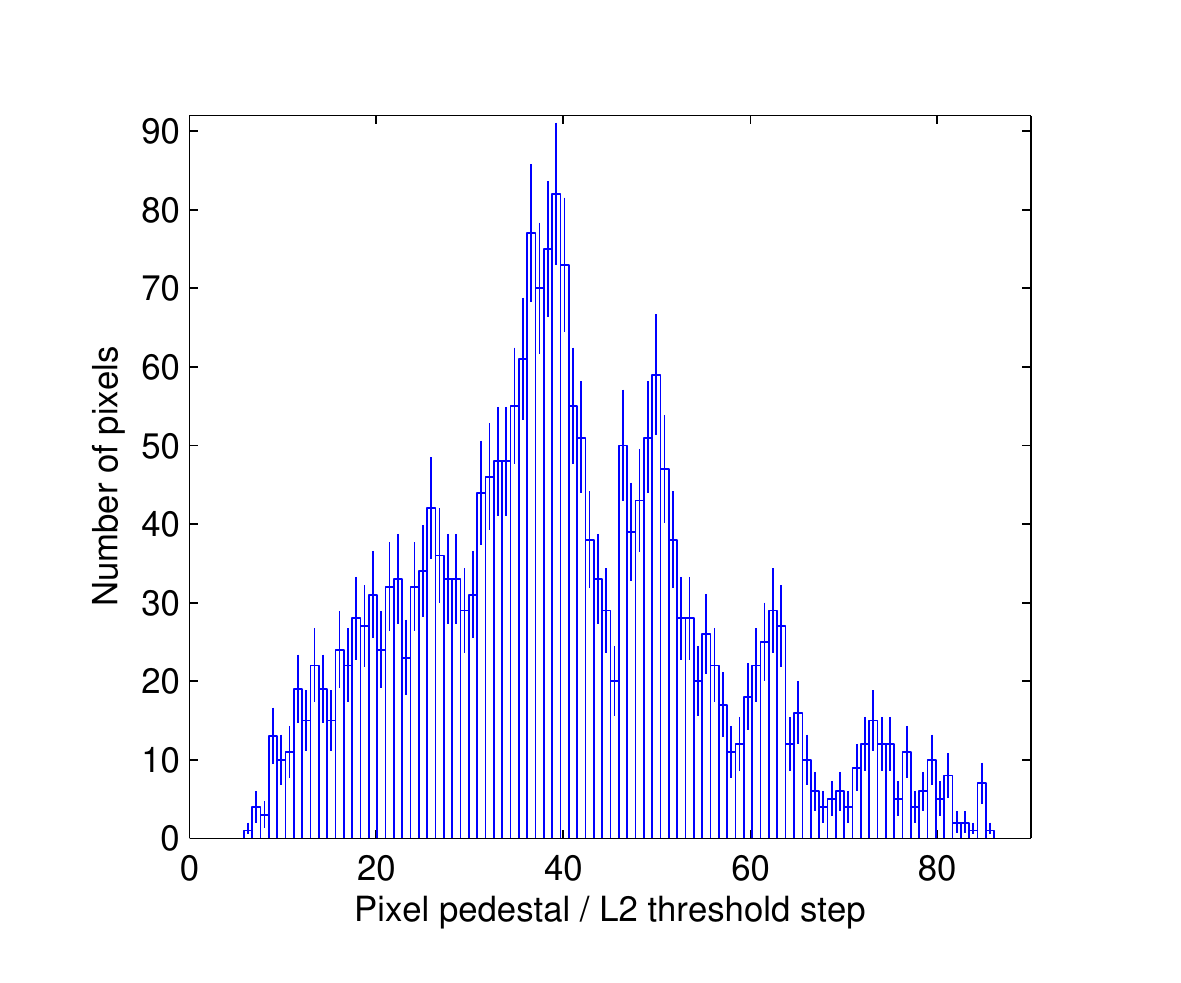}
\end{minipage}
\hspace{0.5cm}
\begin{minipage}[b]{0.5\linewidth}
\centering
\includegraphics[width=\textwidth]{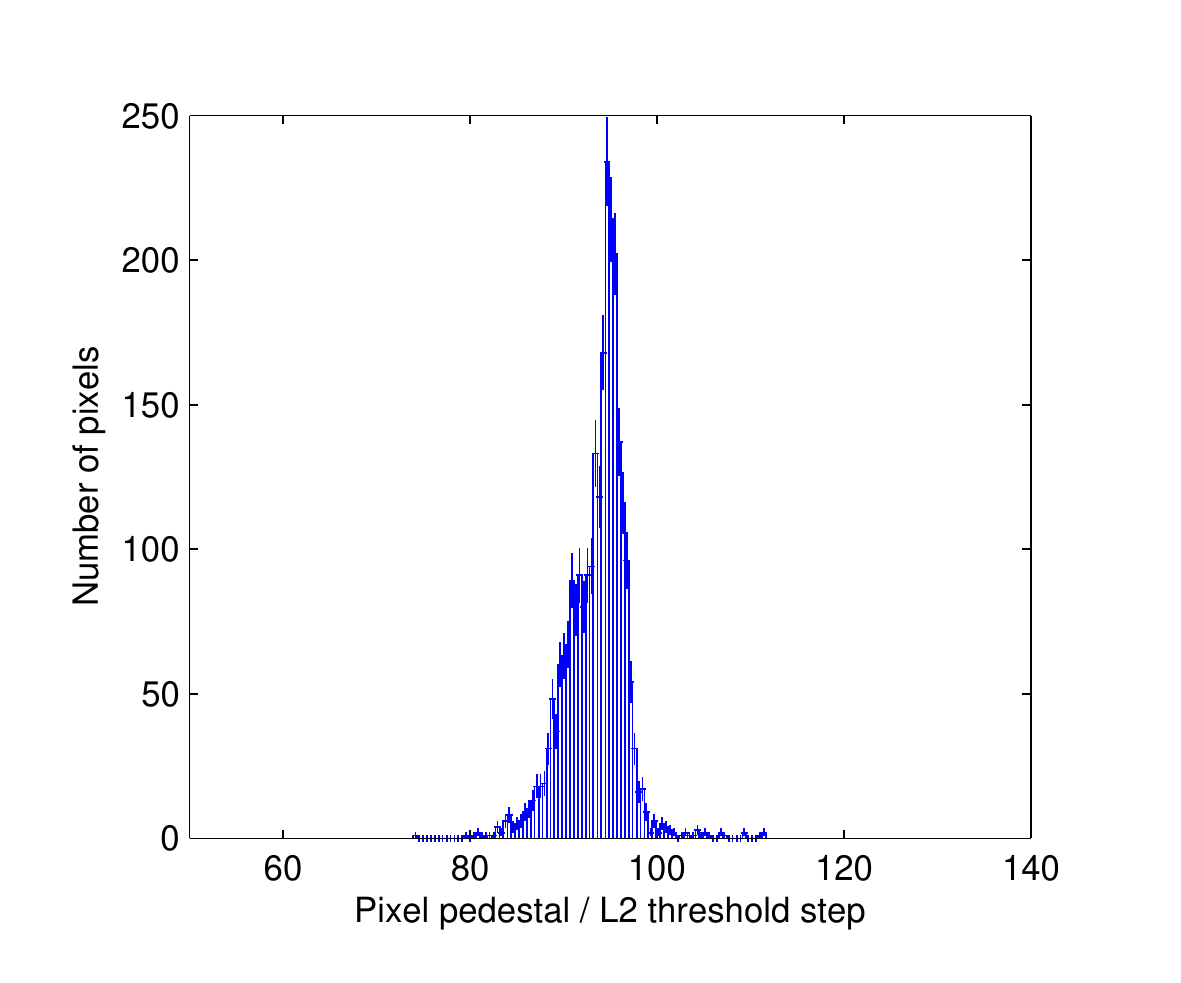}
\end{minipage}
\caption{Distribution of measured pixel pedestals before (left) and after (right) applying comparator offset trims.}
\label{tpo}
\end{figure}

Figure~\ref{tpo} shows the distribution of the extracted pixel pedestals prior to applying comparator offset trims: a large offset range of $\sim$80~threshold units exists. Note that these results are given in terms of the step size of the second threshold level (L2) and that a provisional calibration of these values is given in Section~\ref{Fe55_sec}. This large and irregular pedestal spread results from process variations, transistor threshold offsets, and the differing pixel locations within the column and array. Readout lines from some columns are longer, due to the circuit layout, causing differences in capacitance and voltage drop. These aspects combine to produce the particular pedestal measured for each pixel.

After calibrating the comparator trim, optimal trims were chosen to equalise all offsets, thereby reducing the spread of the measured pixel pedestals. The resultant distribution is also shown in Fig.~\ref{tpo}, demonstrating a greatly reduced spread, with a standard deviation of 3.0~units. This is the minimum spread which could be achieved with the 8-bit comparator trim available. Half-columns of 25 pixels share a single trim value meaning that some residual spread remains, due to pedestal differences within these pixel groups. The skew in the distribution after trimming is due to small differences in the pedestal distributions of the two halves of the array.

\subsection{Pixel noise distribution}

\begin{center}
\begin{figure}[h]
\begin{center}
\includegraphics[width=0.6\textwidth]{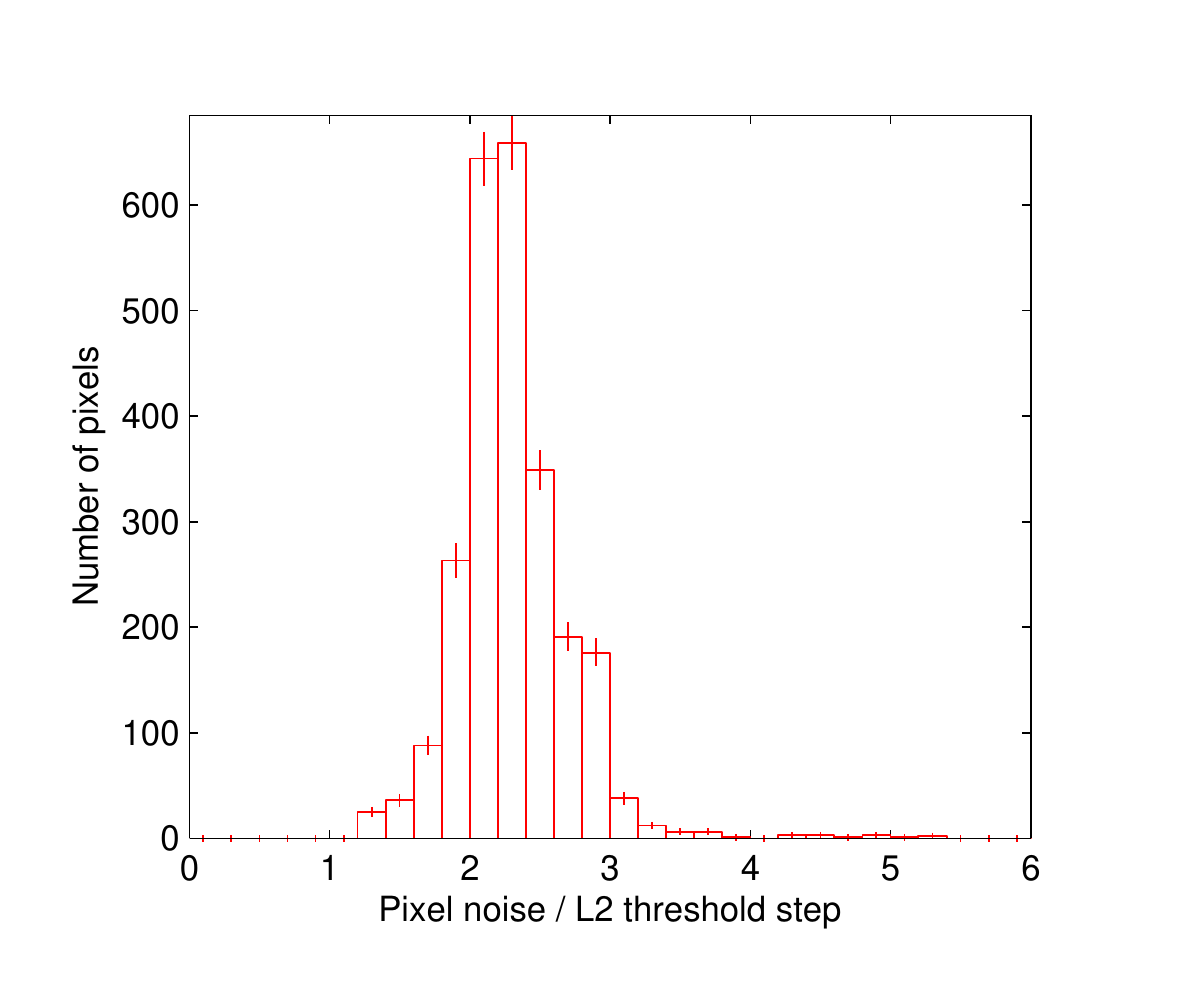}
\caption{Distribution of measured pixel noise values.}
\label{noises}
\end{center}
\end{figure}
\end{center}

The width of the threshold scan curves, approximated as the width of the Error function fit, is a measure of the pixel noise in the absence of radiation. The distribution of these pixel noises is presented in Fig.~\ref{noises}; the most probable value is 2.2~units, with a distribution mean of 2.3~units and standard deviation of 0.4~units. A second sensor tested gave similar values of 1.9, 2.0 and 0.4, respectively.

\subsection{Energy calibration} \label{Fe55_sec}

A $\sim$65~hour exposure using a 160~MBq $^{55}$Fe source was used to provide an absolute energy calibration for the comparator level threshold units. The photoelectrons from K$_{\alpha}$ and K$_{\beta}$ X-rays from $^{55}$Mn fluorescence, with energies of 5.90~keV and 6.49~keV respectively, are expected to be fully absorbed in a silicon volume of order 1~\textmu m$^3$. In general, the resulting ionisation signal will be collected by multiple pixels due to charge diffusion and depending on hit location relative to the diode array. In some cases, however, the charge is collected by a single pixel and thus full-energy photon peaks may be observed in the single-pixel (rather than cluster-sum) spectrum.

\begin{center}
\begin{figure}[h]
\begin{center}
\includegraphics[width=0.5\textwidth]{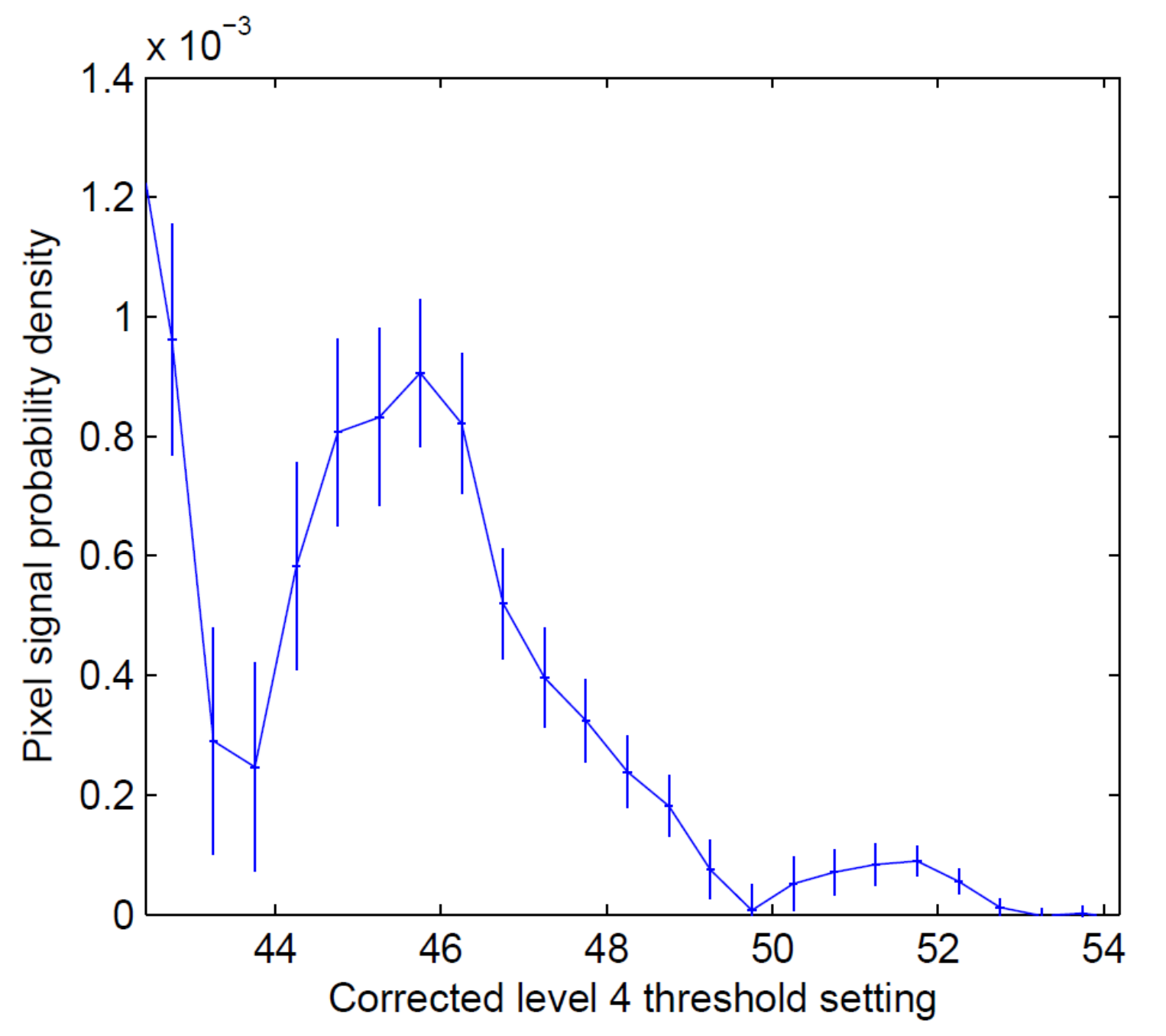}
\caption{Combined pixel spectrum resulting from a 65-hour exposure to $^{55}$Fe. The peaks are identified as corresponding to the full photon energies of 5.90~keV and 6.49~keV. This spectrum is derived from a scan using the fourth threshold level, L4. The horizontal scale has been corrected by subtraction of each pixel pedestal.}
\label{fe55}
\end{center}
\end{figure}
\end{center}

The pixel spectrum obtained is shown in Fig.~\ref{fe55} --- the peaks near 46 and 52 are identified as the full-energy peaks at 5.90~keV and 6.49~keV respectively. This provides a conversion factor of 35.2~e$^-$ for the level~4 threshold step size and thus a step of 8.8~e$^-$ for level 2 (L2\textfractionsolidus L4~=~0.25). This result calibrates the most probable pixel noise as 19.4~e$^-$ and the pedestal dispersion as 26.4~e$^-$. These noise and pedestal values are expected to be sufficient for correct operation of the monitor.

\subsection{Preliminary tests with charged particles}

\begin{center}
\begin{figure}[h]
\begin{center}
\includegraphics[width=0.8\textwidth]{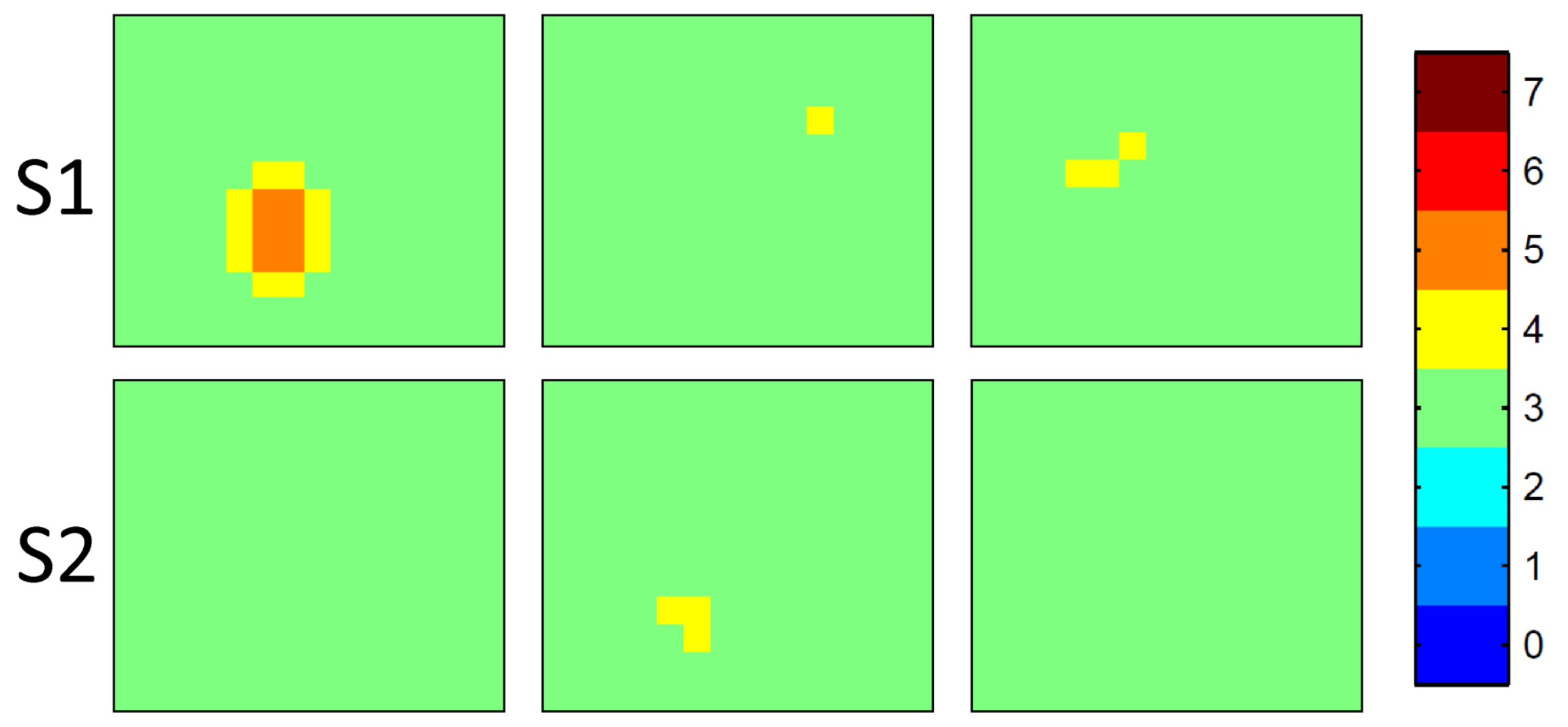}
\caption{Pixel clusters measured during exposure to radioactive sources. The colour scale represents the pixel digital number defined by the set of seven comparator threshold levels. From left to right: a $\sim$5~MeV alpha particle cluster from $^{241}$Am; coincident hit on sensors S1 and S2 during $^{90}$Sr exposure; single hit on S1 during $^{90}$Sr exposure. Note that only a 280\texttimes 340~\textmu m$^2$ portion of the 1~mm$^2$ sensor array is shown in each case.}
\label{blobs}
\end{center}
\end{figure}
\end{center}

As the first stage of a more comprehensive calibration programme, a back-to-back pair of sensors (S1 and S2) was exposed to radiation from $^{241}$Am ($\sim$5.5~MeV~$\alpha$ particles, 60~keV~$\gamma$-rays) and $^{90}$Sr (consecutive $\beta$ decays with E$_{max} = 0.55$ and 2.28~MeV). Prior to operating in radiation monitoring mode with a calibrated set of threshold levels, these measurements were used for simple pixel array imaging (imaging is not used in normal operation). Figure~\ref{blobs} shows examples of hit pixel clusters produced by different particle species/energies. The seven threshold levels were chosen with wide spacings to show relative charge collection only, rather than a calibrated measurement. As shown in the first image, $^{241}$Am alpha particles do not penetrate to S2 and create a large cluster of typically 10-20 pixels.

The second image shows a C2 mode coincident event (simultaneous hits on S1 and S2) obtained during $^{90}$Sr exposure. In this event, the larger cluster on S2 indicates a greater energy deposit than that in S1, as expected from dE\textfractionsolidus dx considerations. The presence of a coincident hit between the two sensors implies an electron with energy greater than $\sim$0.3~MeV (to pass through 500~\textmu m silicon provided by the wafers of S1 and S2). The third image, with a hit on S1 only, indicates an electron with lower energy or an oblique incidence angle. These examples demonstrate the fundamental particle discrimination concepts of the HMRM sensor telescope in action.

\newpage

\section{Conclusion}\label{conc_sec}

The widespread use of satellite radiation monitoring allows improvements in design and mission planning (including estimated lifespan) and introduces the possibility of real-time alerting. Due to the dependence of damage capability on the particle species and energy, particle discrimination is a valuable advantage compared to simple dosimetry. The development of a small, accurate instrument suitable for widespread use on satellites in different orbits could therefore open new prospects for radiation detection in space, with immediate application to commercial as well as scientific payloads. 

The HMRM is a significant development step, with very low mass ($\sim$50~g) and modest engineering/integration costs, while providing a useful and versatile range of data products. These include detailed data characterising individual particle events in addition to simple dosimetry. Monte Carlo simulations have shown that certain particle types, such as low energy protons, may be identified reliably from these event characterisations. More generally, particle energy spectra of useful resolution may be derived from HMRM data using a reconstruction method operating externally to the monitor. The HMRM can become the first space radiation monitor to use CMOS APS devices for particle detection, benefiting from their excellent detection efficiency, radiation tolerance and low noise performance. 

Initial results indicate that the ASIC sensors achieve low noise as anticipated; the project is now proceeding to the calibration and testing phase involving particle beam and radioisotope exposure of fully integrated monitors. Meanwhile, a prototype HMRM is being integrated onto the TechDemoSat-1 satellite which is due to be launched in the near future.

Future phases of the programme will address design iteration and spaceflight qualification, with completion predicted for 2015--2016. A successful HMRM development and qualification process may pave the way for widespread use of this novel technology in many commercial and scientific settings.

\acknowledgments

This work was funded by the European Space Agency (T604-001EE $\&$ T704-027EE) and the Science and Technology Facilities Council (UK), including a PhD studentship and STEP award ST/F007027/1 (EM). We are also grateful for institutional support from Imperial College London.

\bibliographystyle{unsrt}
\bibliography{thesis}

\end{document}